
\tolerance=1200
\magnification 1200
\baselineskip=12pt plus 1pt
\parindent=25pt

\font\small=cmr10 at 10truept
\baselineskip=20pt plus 1pt


{\nopagenumbers
\hfill UMN-TH-1102/1992

\hfill JULY 1992

\vskip1.5cm
{\bf
\centerline{A Closed Bianchi I Universe in
String Theory}
}

\vskip1.5cm

\centerline{{\bf Nemanja Kaloper}$~^{* \dagger}$}
{
\small
\baselineskip=12pt plus 1pt
\footnote{}{$^{*}~~$email: kaloper@umnhep.bitnet }
\footnote{}{$^{\dagger}~~$Address after Sep. 1, 1992:
Dept. of Physics,
Univ. of Alberta, Edmonton, Alta T6G 2J1, Canada }
}
\centerline{School of Physics and Astronomy}
\centerline{University of Minnesota}
\centerline{Minneapolis MN 55455}
\vskip1.5cm
\centerline{\bf Abstract}

{
A special Bianchi I universe is constructed in
$~4D~$ string theory.
Geometrically it represents a $~3D~$
anti-de-Sitter
space crossed with a flat
direction whereas in terms of an associated
conformal field theory
it is an extremal case of a gauged WZWN theory
with target the coset
$~SU(1,1) \times R^2 / R~$. Some of its
properties are discussed.
}

\vfil
\eject}

Recently investigations of string theory
have converged on the study of
nontrivial backgrounds, in hope that new
information on the structure of
quantum gravity may be obtained. A method
which has proven particularly
helpful in discovering novel solutions is
the construction of exact conformal
field theories on the world-sheet via
gauging the Wess-Zumino-Witten-Novikov (WZWN)
sigma model defined on a group of appropriate
topology.
Since Witten's demonstration$^{1)}$ that
such approach, in case of the
$~SL(2,R)~$ group on a $~2D~$ space-time,
yields black
hole solutions, a plethora of
new  configurations have been obtained.
These solutions turn out to admit both
black hole (or black string)$^{1-6)}$
and cosmological
interpretations$^{7-11)}$.

In this letter I will consider a new
conformal field theory on the world sheet,
which can be understood as a diagonal
Bianchi I cosmology. It turns out to be
a closed, anisotropic universe, which
recollapses after existing only a finite
time, much like the recently constructed
example of Nappi and Witten$^{10)}$.
They have found a closed recollapsing
universe constructed as a conformal
field theory on the coset
$~SU(1,1) \times SU(2)/ R^2~$ and
investigated
its singularity structure.
However, this solution is free of
curvature singularities,
as its global structure is
that of a $~S^1 \times R^2~$ anti-de-Sitter
space-time crossed with a flat
direction $~R~$. It is plagued with
singularities in its causal structure,
as closed
timelike loops exist in it.
It represents an extremal case of the
gauged WZWN model on
the coset $~SU(1,1) \times R^2 / R~$
and some internal conformal theory, where
gauging is performed with  taking the
limit of the coupling of one of the
free bosons to $~\infty~$ and keeping
the other decoupled. This in effect
decouples the $~SU(1,1)~$ group, too.
Such extremal limit corresponds to the one
discussed by Horne and Horowitz$^{4)}$
in relation to extremal black strings
in $~3D~$. A
peculiarity of the solution is the
constant dilaton, whose evolution is
suppresed by a cancelation between the
Kalb-Ramond charge and the effective
cosmological constant, resulting from
the central charge deficit.

The action defining the effective field
theory on the target space is,
in the world sheet frame
$$
S~=~\int d^{4}x\sqrt{G}  e^{-\sqrt{2}\kappa \Phi}
\big({1 \over 2\kappa^{2}}R -
H_{\mu\nu\lambda}H^{\mu\nu\lambda}+
\partial_{\mu}\Phi \partial^{\mu} \Phi
+ \Lambda \big) \eqno(1)
$$

\noindent  where $~G_{\mu\nu}~$ and
$~H_{\mu\nu\lambda}=\partial_{[\lambda}
B_{\mu\nu]}~$  and $~\Phi~$
are the world sheet target
metric, the Kalb-Ramond antisymmetric
field strength and the dilaton.
Braces denote antisymmetrization over
enclosed indices.
Here the cosmological constant has
been included to represent the
central charge deficit
$~\Lambda={2 \over 3}\delta c_{T}
= {2 \over 3}(c_{T}-4) \ge 0~$. It arises
as the difference of the
internal theory central charge and
the total central charge for a
conformally invariant theory $~ c_{tot}=26~$.
In the case I will consider,
when the sigma model
$$
S_{\sigma}~={1 \over \pi}
{}~\int d^2\sigma (G_{\mu\nu} + \sqrt{2 \over 3}
B_{\mu\nu})~\partial_{+}X^{\mu}
\partial_{-}X^{\nu}
\eqno(2)
$$

\noindent is defined on the coset
$~SU(1,1) \times R^2 / R~$, the central
charge of the target for level $~k~$ is
$~c_{T}=(3k / k-2) + 2 - 1~$,
where $~2~$ and $~-1~$ correspond to the
two free bosons and gauging,
respectively. The $\sqrt{2/3}$ in Eq. (2)
is introduced following the
normalization adopted in (1).
The central charge deficit, by the formulae
above, will be given by
$$
\delta c_{T}~=~{6 \over k - 2}~\simeq ~{6 \over k}
\eqno(3)
$$

\noindent in the semiclassical limit
$~k \rightarrow \infty~$,
where the theory is most reliable.

I will first consider the effective
field theory as described by the action (1)
and search for a specific class of
solutions featuring toroidal symmetry and
constant dilaton. Such solutions
were noticed in [ 4 ] as a special class of
extremal black strings in $~3D~$.
So, for the background metric of the
Bianchi I type,
$$
ds^{2}=-e^{2\nu}dt^{2}+e^{2\lambda}dx^{2}
+e^{2\mu}dy^{2}+e^{2\eta}dz^{2}
\eqno(4)
$$

\noindent where the metric functions depend
only on time, it is easy to
derive beta function equations from (1).
The axion equations
$~\partial_{[\mu}H_{\nu\lambda\sigma]}=0~$
and
$~\partial_{\mu}\Bigl(\sqrt{G}\exp(-\sqrt{2}\Phi)
H^{\mu\lambda\sigma}\Big)=0~$
are solved by (the topological charge term)
$$\eqalign{
B~=&~2Y~dx \wedge dy \cr
\dot Y~=~-{Q \over 2}
\exp(&\sqrt{2}\Phi + \nu + \lambda + \mu - \eta)
\cr}
\eqno(5)
$$

\noindent where the overdot denotes the
time derivative.
With the choice of gauge
$~\nu=\lambda + \mu~$ and after enforcing
$~\Phi=\Phi_0=const~$ (which implies
$~\eta=const~$ and by rescaling the $z$
coordinate it can be set equal to zero),
they reduce to
$$\eqalign{
\Lambda~&=~{Q^2 \over 3}
e^{2\sqrt{2} \Phi_0} \cr
\ddot \mu~=&~\ddot \lambda~
=~-\Lambda e^{2(\lambda + \mu )}\cr
2 \dot \lambda \dot \mu &~
=~-\Lambda e^{2(\lambda + \mu )} \cr}
\eqno(6)
$$

\noindent which are very easy to solve.
The solution is,
$$\eqalign{
ds^2~&=~-{\xi^2 \over 2 \Lambda}\,
{dt^2 \over \cosh^2\xi t}
+ { e^{2\lambda_0 + \xi t} dx^2
+ e^{2\mu_0 - \xi t} dy^2 \over \cosh\xi t }
+ dz^2  \cr
&~~~~~~B~=~\bigl(B_0 -
\sqrt{3 \over 4 \Lambda}\, \xi \tanh{\xi t} \,\bigr)
{}~dx \wedge dy   \cr}
\eqno(7)
$$

\noindent with integration constants
related by
$~Q^2 = 3\Lambda \exp{(-2\sqrt{2}\Phi_0 )}~$ and
$~\exp(\lambda_0 + \mu_0) = \xi / \sqrt{2\Lambda}~$.
After a coordinate transformation
$~\tau = \tanh \xi t~$,
$~x \rightarrow \exp(-\lambda_0) x~$
$~y \rightarrow \exp(-\mu_0) y~$,
which imposes $~-1 \le \tau \le 1~$,
the solution becomes
$$\eqalign{
ds^2~&=~-{1 \over 2 \Lambda}\,
{d\tau^2 \over 1 - \tau^2} + (1 + \tau) dx^2
+ (1 - \tau) dy^2 + dz^2 \cr
&~~~~~~~~~~B~=~\bigl(B_0 -
\sqrt{3 \over 2 }\, \tau \,\bigr) ~dx \wedge dy  \cr}
\eqno(8)
$$

\noindent This solution apparently
can be extended to
describe one cosmological sector
($~-1 \le \tau \le 1~$) and two ``static''
string-like branches ($~\tau \le -1~$
and $~\tau \ge 1~$). This is the case
discussed in Ref. [4-6] where the group
was $~SL(2,R)~$.
However, a future oriented
observer inside the cosmological patch,
starting from anywhere with
$~\mid \tau \mid < 1~$ can not
enter the future ``static'' branch,
which is clear since $~\tau=1~$ is a
well-defined horizon. Therefore,
this observer is
confined within the cosmological
sector throughout its life, and
it is in this way how he/she perceives
the solution.
I will hence concentrate on the
cosmological sector.
An additional coordinate
transformation $~\tau=\sin \theta~$ recasts the
metric to
$$
ds^2~=~-{1 \over 2 \Lambda}\,
d\theta^2 + (1 + \sin \theta) dx^2
+ (1 - \sin \theta) dy^2 + dz^2
\eqno(9)
$$

\noindent where $~\theta \in (-\pi /2, \pi /2)~$
by the constraints on $~\tau~$
(I choose the principal branch to define
$~\theta~$).
This form of the metric suggests another
possibility to extend the solution.
First, one should note that a
universe described by
the metric above undergoes expansion
from zero volume (at $~\theta=-\pi /2~$)
through maximum size (at $~\theta=0~$)
back to collapse (at $~\theta=\pi /2~$).
A simple calculation shows that
$~R_{3 \mu\nu\lambda}=0~$ and that the
restriction to a $~3D~$ slice $~z=const~$
is a maximally symmetric
anti-de-Sitter space-time with
$~R_{\mu\nu\lambda\sigma}
=-(\Lambda /2)(g_{\mu\lambda}
g_{\nu\sigma}-g_{\mu\sigma}g_{\nu\lambda})~$.
Hence curvature singularities do not appear.
However, if the time coordinate is
compactified to a circle $~S^1~$, time-like
loops appear, and singularities
plague causal structure of the space-time.
This apparent violation of the cosmic
censorship conjecture, appearing through
the (hidden) singular behavior
and reduction of the
metric to three dimensions at
$~\theta=\pm \pi /2~$
is remedied by the fact that the
whole universe  collapses then,
as volume vanishes, just as discussed by
Nappi and
Witten$^{10)}$. Yet, at the first sight
one might
think that it is possible to extend the
solution past $~\theta=\pm \pi /2~$ and
obtain a solution representing an oscillating
cosmos, perhaps even avoiding the
closed time-like lines.

This dilemma is answered properly
by the WZWN construction.
The resolution of this
question is a demonstration of the
power of the WZWN constructions to resolve
ambiguities such as the one above.
It firmly states that by the construction,
the time coordinate must live on a circle.

The WZWN approach starts with a choice of
a group manifold $~G~$ on which
sigma model on level $~k~$ is
$$\eqalign{
S_{\sigma}~=~{k \over 4 \pi} \int d^2\sigma
Tr\bigl(g^{-1}\partial_{+}g\,&g^{-1}
\partial_{-}g\Bigr)
- {k \over 12 \pi}\,\Gamma(g)\cr
\Gamma(g)~=~\int_{M} d^3 \zeta
Tr\Bigl( g^{-1}dg &\wedge g^{-1}dg
\wedge g^{-1}dg \Bigr) \cr}
\eqno(10)
$$

\noindent The action is gauged with
introduction of gauge
fields which couple to
an anomaly-free subgroup of the
invariance group $~G \times G~$ of
the sigma model.
It is well known that the only
anomaly-free subgroups
are vector, axial vector and
chiral subgroups. I will choose
the axial gauging of the theory,
starting with the
group manifold $~SU(1,1) \times R^2~$
and then factor out the one-parameter
axial invariance subgroup
of the target $~SU(1,1)~$ mixed with
translations along one of the free bosons.
This choice is similar to that of
Refs. [ 4,5 ]. The only difference is the
global structure of the group $~SU(1,1)~$
here versus theirs $~SL(2,R)~$.
Since the group $~SU(1,1)~$ is locally
given by
$$
 g~=~\left(\matrix{~a~&~u~\cr
 -v~&~b~\cr}\right)
\eqno(11)
$$

\noindent the explicit form
of the chosen ungauged sigma model is
$$\eqalign{
S_{\sigma}~=~&-{k \over 4 \pi}
\int d^2\sigma \Bigl(\partial_{+}u \partial_{-}v
+ \partial_{-}u \partial_{+}v +
\partial_{+}a \partial_{-}b
 + \partial_{-}a \partial_{+}b \Bigr) \cr
&+ {k \over 2 \pi} \int d^2\sigma
\ln{u} \Bigl(\partial_{+}a \partial_{-}b
- \partial_{-}a \partial_{+}b \Bigr)
+ {1 \over \pi} \int d^2\sigma \sum_j
\Bigl(\partial_{+}x_j \partial_{-}x_j \Bigr) \cr}
\eqno(12)
$$

\noindent The gauge transformations are
$$\eqalign{
&\delta a = 2\epsilon a ~~~~~\delta b
= 2\epsilon b
{}~~~~~\delta u = \delta v = 0 \cr
&\delta x_{1} = 2\epsilon c ~~~~~
\delta x_{2} = 0
{}~~~~~\delta A_{j} = -\partial_{j}
\epsilon \cr}
\eqno(13)
$$

\noindent and the gauged sigma model (9) is
$$\eqalign{
S_{\sigma}(g, A)~=~S_{\sigma}(g)
&+ {k \over 2 \pi} \int d^2\sigma A_{+}
\Bigl( b\partial_{-}a - a\partial_{-}b
- u\partial_{-}v + v\partial_{-}u +
{4 c \over k} \partial_{-}x_{1} \Bigr) \cr
&+ {k \over 2 \pi} \int d^2\sigma A_{-}
\Bigl( b\partial_{+}a - a\partial_{+}b
- v\partial_{+}u + u\partial_{+}v +
{4 c \over k} \partial_{+}x_{1} \Bigr) \cr
&+ {k \over 2 \pi} \int d^2\sigma 4 A_{+}A_{-}
\Bigl( 1 + {2 c^2 \over k} - uv \Bigr)
\cr}
\eqno(14)
$$

There is no $~A_{\pm}~$ dependent contributions
from second term from (10) in (14)
since the  gauge group is anomaly-free.
The next step is to integrate out the gauge
fields, fix the gauge of the group (by
choosing $~a=\pm b~$ to eliminate the
unpleasant logarythm from (12), and picking
the sign so that still $~\det{g}=1~$)
rescale $~x_1 \rightarrow (2c/ \sqrt{k}) x_1~$ and
take the limit $~c \rightarrow \infty~$
which effectively decouples
the  $~SU(1,1)~$ part from the gauge fields.
In the end, the resulting action can
be rewritten as
$$\eqalign{
S_{\sigma ~eff}~=~&-{k \over 8 \pi}
\int d^2\sigma
{v^2 \partial_{+}u \partial_{-}u + u^2
\partial_{-}v \partial_{+}v +
(2 - uv)\bigl(\partial_{+}u \partial_{-}v
+ \partial_{-}u \partial_{+}v \bigr)
\over (1 - uv)} \cr
&+ {1 \over \pi} \int d^2\sigma \Bigl(
2 (1 - uv) \partial_{+}x_1 \partial_{-}x_1
+ \partial_{+}x_2 \partial_{-}x_2 \Bigr) \cr
&+ {\sqrt{k} \over 2\pi} \int d^2\sigma
\Bigl( \bigl( u\partial_{-}v
- v\partial_{-}u \bigr) \partial_{+}x_1 +
\bigl( v\partial_{+}u
- u\partial_{+}v \bigr) \partial_{-}x_1 \Bigr)
\cr}
\eqno(15)
$$

\noindent A transformation of coordinates
covering the cosmological sector of the
model is, with $~\theta \in (-\pi / 2, \pi / 2)~$,
$$
u~=~i~{e^{-{2\over \sqrt{k}}y}
\over \sqrt{2}} \sqrt{1 - \sin \theta}
{}~~~~~~~~~~~~~~
v~=~-i~{e^{{2 \over \sqrt{k}}y}
\over \sqrt{2}} \sqrt{1 - \sin \theta}
\eqno(16)
$$

\noindent which differs from the
one employed in
Ref. [ 4,5 ] by where it is valid
in the parameter space, and thence
lends to the above cosmological
interpretation of the solution.
Relabeling $~x_1=x,~x_2=z~$ and
comparing with the sigma model
action (2) yields the answer for the metric
and the axion:
$$\eqalign{
& ~~~~~~~G_{\mu\nu}~=~
\left(\matrix{-{k \over 8}
&~0~&~0~&~0~\cr
{}~0~&1 + \sin \theta&~0~&~0~\cr
{}~0~&~0~&1 - \sin \theta&~0~\cr
{}~0~&~0~&~0~&~1~\cr}\right) \cr
& B_{\mu\nu}~=~\left(\matrix{
{}~0~&~0~&~0~&~0~\cr
{}~0~&~0~&\sqrt{3 \over 2}
(1 - \sin \theta)&~0~\cr
{}~0~& - \sqrt{3 \over 2}
(1 - \sin \theta)&~0~&~0~\cr
{}~0~&~0~&~0~&~0~\cr}\right) \cr}
\eqno(17)
$$

\noindent The dilaton can be found
either from a careful computation of the
Jacobian determinant arising from
integrating out the gauge fields, or from the
associated  effective action. This
has already been done in the ans\"atz
preceeding Eq. (6), where it was
a constant. Indeed, inspection of the Jacobian
matrix before the limit
$~c \rightarrow \infty~$ is taken shows
that it is $~\propto 1/(1 + (2c^2 /k) - uv)~$
$= (k/2c^2)/\bigl(1 + (k/2c^2)(1 - uv)\bigr)~$
(see Ref. [ 3 ]).
As $~c \rightarrow \infty~$ the non-constant
terms decouple and do not contribute to the
dilaton. The final answer for the
metric indeed is Eq. (9), after using
$~\Lambda=4 / k~$. As a result of this
construction, it is clear that the time
$~\theta~$ lives on a circle, since
$~\theta'=\theta + n\pi~$ represent the same
group element of $~SU(1,1)~$. Hence
global structure of the manifold is
$~S^1 \times R^3~$ and continuation of
$~\theta~$ beyond the interval
$~(-\pi/2, \pi /2)~$ inevitably leads to the
appearance of time-like loops and
singularities in the causal structure of the
space-time. Notice that the point
$~\theta=-\pi /2~$ ($~\tau=-1~$)
where the gauge
fixing breaks down corresponds to the
``Big Bang'' in this model, and as it was
argued before, may not be a reliable
point of the theory. In other words, it is
expected$^{12)}$ that a different (topological)
field theory should describe the model in the
vicinity of $~\theta=-\pi /2~$ and account
for effects of quantum gravity. This is
underlined even more with the
above cosmological interpretation.

In summary, I have considered a gauged
WZWN model on the coset
$~SU(1,1) \times R^2 /R~$ and have
demonstrated that a sector of the conformal
theory admits a cosmological interpretation.
The universe described by it is a $~3D~$
anti-de-Sitter manifold crossed with a
flat direction. It evolves as a closed universe,
starting from a ``Big Bang'' and reaching a
``Big Crunch'' after a finite comoving
time. The ``Big Crunch'' is in some
sense welcome in the model since it prevents
appearance of causal singularities
which would exist if time is to be extended
beyond the moments of creation and annihilation.

\vskip1cm
{\bf Acknowledgements}
\vskip0.5cm

This work has been supported in part by
the University of Minnesota Doctoral
Dissertation Fellowship.

\vskip1cm
{\bf References}
\vskip0.5cm
\item{[1]} E. Witten, Phys. Rev. {\bf D44} (1991) 314.
\item{[2]} N. Ishibashi, M. Li and A.R. Steif,
Phys. Rev. Lett. {\bf 67}
(1991) 3336;
P. Ginsparg and F. Quevedo,
Los Alamos preprint LA-UR-92-640,
Feb. 1992;
I. Bars and K. Sfetsos,
USC preprint USC-92/HEP-B1, May 1992;
I. Bars and K. Sfetsos,
USC preprint USC-92/HEP-B2, May 1992;
K. Sfetsos, USC preprint USC-92/HEP-K1, June 1992.
\item{[3]} P. Horava,
Phys. Lett. {\bf B278} (1992) 101;
E.B. Kiritisis,
Mod. Phys. Lett. {\bf A6} (1991) 2871;
D. Gershon,
Tel Aviv University preprint TAUP-1937-91, Dec 1991.
\item{[4]} J.H. Horne and G.T. Horowitz,
Nucl. Phys. {\bf B368} (1992) 444;
\item{[5]} E. Raiten,
Fermilab preprint FERMILAB-PUB-91-338-T, Dec. 1991.
\item{[6]} N. Kaloper,
Univ. of Minnesota preprint UMN-TH-1024/92, May 1992.
\item{[7]} J. Antoniades, C. Bachas, J. Ellis
and D. Nanopoulos,
Phys. Lett. {\bf B221} (1988) {393;
J. Antoniades, C. Bachas and A. Sagnotti,
Phys. Let. {\bf B235} (1990) 255.
\item{[8]} K. Behrndt,
DESY preprint DESY-92-055-REV, April 1992.
\item{[9]} C. Kounnas and D. Lust,
CERN preprint CERN-TH-6494-92, June 1992.
\item{[10]} C.R. Nappi and E. Witten,
IAS preprint IASSNS-HEP-92/38, June 1992.
\item{[11]} for a review see
A.A. Tseytlin, Cambridge
University preprint, DAMTP-92-06, June 92.
\item{[12]} T. Eguchi,
Univ. of Chicago preprint EFI-91-58, Oct. 1991.
\bye